\documentclass{JHEP3}
\usepackage{epsfig,amsmath,amsthm}
\newcommand{\be}{\begin{equation}}
\newcommand{\ee}{\end{equation}}
\newcommand{\bea}{\begin{eqnarray}}
\newcommand{\eea}{\end{eqnarray}}
\def\bse{\begin{subequations}}
\def\ese{\end{subequations}}

\def\IZ{\relax\ifmmode\hbox{Z\kern-.4em Z}\else{Z\kern-.4em Z}\fi}

\newcommand{\non}{\nonumber \\}

\def\cl{{\cal L}} \def\co{{\cal O}}

\def\tphi{\widetilde{\phi}}

\def\al{\alpha} \def\bt{\beta}
  \def\eps{\epsilon}

\def\presub{\vspace{.5cm} \noindent}

\def\bi{\begin{itemize}} \def\ei{\end{itemize}}

\def\Schw{Schwarzschild }
\def\({\left(} \def\){\right)}
\def\[{\left[} \def\]{\right]}
\title{ \center{Higher Order Perturbations Around Backgrounds\\ with One Non-Homogeneous Dimension}}

\author{
Barak Kol \\
 Racah Institute of Physics\\
 Hebrew University \\
 Jerusalem 91904,
 Israel\\
{\tt barak\_kol@phys.huji.ac.il}}

\abstract{It is shown that perturbations around backgrounds with
one non-homogeneous dimension, namely of co-homogeneity 1, can be
canonically simplified, a property that is shown to hold to any
order in perturbation theory. Recalling that the problem naturally
reduces to 1d, a procedure is described whereby for each gauge
function in 1d two 1d fields are eliminated from the action -- one
is gauge and can be eliminated without a constraint and the other
is auxiliary. These results generalize the results of
hep-th/0609001 from linear to non-linear perturbations and they
unify two cases of physical interest: cosmological perturbations
and perturbations to static spherically symmetric backgrounds. An
application to black strings is discussed in some detail.}

\begin{document}

\section{Introduction}

Analysis of perturbations (gravitational waves) around a given
background in General Relativity, and in particular making a
judicious choice of gauge, is a problem whose complexity grows
with the degree of co-homogeneity of the background, namely the
number of non-homogeneous coordinates. For co-homogeneity zero, we
are dealing with the well-understood case of a maximally symmetric
space time (flat space, de-Sitter or Anti de-Sitter) and once
harmonic analysis is employed the problem reduces altogether to
algebra. The case of co-homogeneity one was solved at the linear
level in \cite{1dPert} following a concrete analysis of the black
hole negative mode \cite{nGPY} thereby unifying known results in
two cases of physical interest: static spherically symmetric
space-times (mostly black holes)
\cite{Moncrief1974,GerlachSengupta} and spatially homogeneous
cosmologies \cite{Bardeen,MukhanovFeldmanBrandenberger}. The
solution for co-homogeneity two and higher is unknown and one may
expect it to be challenging given the complexity of a specific
case, that of perturbations around the Kerr spacetime (where
Newman-Penrose variables are employed).

Let us briefly review the results of \cite{1dPert}. It was shown
that once the natural reduction to a 1d theory is carried out, for
each 1d gauge function two 1d fields can be eliminated: one
through the gauge and the other being auxiliary. The
transformation to the new set of fields is local in the
non-homogeneous coordinate and invertible. This procedure proves
that in 1d the gauge can be completely eliminated and an
invertible transformation exists into gauge-invariant fields,
which is presumably the reason that the cosmological version of
this procedure is known as ``gauge invariant perturbation
theory''. One should stress that while it is trivially true that
all physical observables including perturbations must be gauge
invariant it is certainly not true in general that one can
transform (one to one and locally) into gauge-invariant fields --
this property holds only for co-homogeneity one (or zero).
Actually, one can work in the action formulation and perform a
\emph{constraint free gauge-fix} of certain fields (the
constraints are contained already in the equations of motion).
Moreover, not only can the gauge be eliminated thereby eliminating
a similar number of fields, but also each 1d gauge function is
responsible for a 1d auxiliary field which helps to further
decouple the action. Accordingly it is said that ``the gauge
shoots twice''. This auxiliary field is sometimes considered to be
a constraint, though from the current perspective it is certainly
not one: a constraint is a residual equation of motion left after
fixing the gauge while here the gauge gets eliminated altogether.
% residual freedom to choose the gauge. Covariant perturbations?

The results of \cite{1dPert} were applied to obtain new insight
into gravitational waves in the \Schw background (Regge-Wheeler
and Zerilli equations) \cite{nRW} and to a study of the dependence
of the black hole negative mode on the space-time dimension
\cite{high&lowD}.

One wonders whether the features of linear perturbations carry
over to the non-linear case and in particular whether each 1d
gauge function can be continued to be used to completely eliminate
two 1d fields. This question on non-linear perturbations is
further motivated by applications including cosmology as well as
for the computation of the order of phase transition associated
with the Gregory-Laflamme black string instability
\cite{Gubser,SorkinD*,LG-GL}.

Experience suggests that the linear part of the equations sets
much of their qualitative properties. Indeed, in section
\ref{general-sect} we provide and justify a procedure that answers
affirmatively and to all orders the questions from the preceding
paragraph. This work can be thought to unify the existing
literature on non-linear cosmological perturbations
\cite{Nakamura,RigopoulosShellard,LythMalikSasaki,LangloisVernizzi}
with that of non-linear perturbations in spherical symmetry
\cite{Brizuela-etal}. In section \ref{apply-sect} we spell out in
greater detail the application of the general procedure to the
non-uniform string, but we leave its implementation to future
work.

\presub \emph{Comment}: This paper is not intended to be submitted
to a refereed journal. Papers such as this, which suggest a new
computational procedure or theory, are easier to referee once they
are applied and demonstrated in full rather than as an abstract
argument, since the former confirm the paper's validity and
usefulness beyond doubt. Indeed, I plan to apply the procedure
presented here to the case of the non-uniform black string. On the
other hand, it is plain that a general procedure together with its
rational are self-contained and have a value on their own that
merits their publication. Thus in order to avoid unnecessary
difficulties in the process of refereeing I choose to post this
paper on the archives, but postpone submitting it to a refereed
journal until it can be strengthened by an application.

\section{The procedure}
\label{general-sect}

We start this section by stating a canonical procedure to handle
non-linear 1d actions with gauge, and then we continue to explain
and justify it. The procedure is \bi
 \item Identify the fields and their gauging and thereby the derivatively-gauged (DG) fields, namely 1d tensors
 or vectors.
 \item It is more economical to fix a gauge before obtaining the
 action. For each gauge function we may eliminate any single field (in
 whose variation the gauge function appears), but not a DG
 field. This is a constraint-free gauge-fixing \cite{1dPert}.
 \item Obtain the action (up to the required order in the fields).
 \item Write the equations of motion for the DG fields.
 \item Use the equations of motion to solve for the DG fields. The
 solution is obtained through series inversion and should be expanded up to
 the required order.
 \item Substitute the expression for the DG fields back into the
 action.

 Alternatively, one may skip the gauge-fixing step, and upon substituting the DG
 fields the action will be found to depend only on certain
 gauge-invariant fields.
 \ei

\vspace{.5cm}

 Let us now discuss each item in detail.

\noindent \emph{Fields and general set-up}. After separation of
variables, namely expanding all fields in harmonic functions of
the homogeneous coordinates, we may reduce the perturbation
problem to 1d by performing the integration over the homogeneous
coordinates. More generally we consider any 1d action with a gauge
symmetry. We denote the 1d variable \footnote{In the cosmological
set-up $x$ would normally be denoted by $t$, while in spherically
symmetric backgrounds it would normally be denoted by $r$.}
 by $x$ and the fields and gauge fields respectively by the vector
notation \bea
 \phi&=&\phi^i(x) \qquad i=1,\dots,n_F \non
 \xi&=&\xi^a(x) \qquad a=1,\dots,n_G \eea
where $n_F,\, n_G$ are the number of (real) fields and gauge
functions, respectively.

\presub \emph{Identifying the derivatively-gauged (DG) fields}.
Let us recall the definition of the DG fields \cite{1dPert}. We
assume the gauging to be linear in the gauge functions and at most
linear in derivatives, which is indeed the case when analyzing
perturbations in GR or gauge theory. The gauge variation is given
by \be
 \delta \phi = G_1\, \xi' + G_0\, \xi \ee
where $\(G_1(x),\,G_0(x,\phi,\phi') \)$ is a pair of $n_F \times
n_G$ matrices which depend on the coordinate $x$. We allow $G_0$
to depend also on the fields $\phi$ and their derivatives
(non-linear contributions to the gauging), but we assume $G_1$ to
have no such dependence, which is indeed the case in GR
perturbations.

\noindent \emph{Definition}: The image of $G_1$ in field space is
called {\it the derivatively-gauged (DG) fields}.

We recognize this algebraic definition to correspond to the
standard gauge variation of fields which are vectors or tensors
from the 1d point of view.

Following the identification of the DG fields, the field space can
be split into the DG-subspace with coordinates $\phi_{DG}^\al$ and
an arbitrarily chosen complement with coordinates $\phi_X^r$,
namely \be
 \phi^i=(\phi_{DG}^\al,\, \phi_X^r) ~~~~\al=1,\dots,n_G\, , ~~ r=1,\dots,n_F-n_G. \ee

\presub \emph{ Obtaining the action}. The action \be
 S = S[\phi] \ee
is considered as the input to the procedure. In practice one needs
to reduce the perturbation problem to 1d as already mentioned. To
simplify the computation a constraint-free gauge-fix should be
performed by fixing for each gauge function any field (in whose
variation the gauge function appears), as long as it is not a DG
field.

\presub \emph{ Writing the equations of motion for the DG fields}.
This is a straightforward variation of the action.

\presub \emph{ Solving for the DG-fields order by order and substituting back into the action}. The
equations of motion for the DG fields read \be
 0= \frac{\delta S}{\delta \phi_{DG}} = L_{DG}\, \phi_{DG} + \cl_m\(\phi_X\)+ \co \(
 \phi_X,\phi_{DG} \)^2 \label{DGEOM} ~. \ee
This expression highlights the linear part of the equations as
obtained in \cite{1dPert}. We suppressed the indices, $L_{DG}$ is
just a $x$-dependent $n_G \times n_G$ matrix which we assume to be
invertible, and $\cl_m$ is a linear operator with
 at most one derivative which mixes $\phi_X$ into the equations of motion of $\phi_{DG}$.

We proceed to perform a change of variables from $\phi_{DG}$ to $\tphi_{DG}$
and substitute it back into the action. However, in practice we shall find that
it suffices to substitute zero for $\tphi_{DG}$ thereby somewhat simplifying the
procedure. Accordingly we start by explaining the change of variables, but for
 practical purposes one may wish to skip to the paragraph of
 (\ref{phiDG-soln}).

We define \be
 \tphi_{DG}:=L_{DG}^{-1}\, \frac{\delta S}{\delta
\phi_{DG}} = \phi_{DG}  + L_{DG}^{-1}\, \cl_m\(\phi_X\)+ \co \(
 \phi_X,\phi_{DG} \)^2 \label{def-tphi} \ee where we used (\ref{DGEOM}) in the second equality to recognize
 that at the linear level $\tphi_{DG}$ is a shifted version of $\phi_{DG}$.

In order to substitute the change of variables into the action we need to invert
this relation and obtain $\phi_{DG}=\phi_{DG}(\tphi_{DG},\phi_X)$. Since we are dealing
 with a perturbation theory it suffices to perform the inversion perturbatively, and since
$L_{DG}$ is invertible that can always be done through a standard procedure as we shall
detail shortly.

Next we substitute the inverted relation into the action. Note
that the determinant of the transformation is no longer unity, as
was the case for linear perturbations, but there is no need to
account for that as long as classical physics is involved.
Actually by construction (\ref{def-tphi}) the equations of motion
state that \be \tphi_{DG}=0  \label{tphi0} \ee
 and hence this  sector is guaranteed to decouple,
and it suffices to substitute $\tphi_{DG} \to 0$. Note that equations (\ref{tphi0})
can be interpreted to define a natural
 complement in field space to the DG fields, rather than the
 arbitrarily complement $X$ chosen before. Moreover, since zero is
 gauge-invariant so are the $\tphi_{DG}$ fields.

Having realized that $\tphi$ can be put to zero, let us describe how this
 could be implemented in practice to avoid defining $\tphi$ in the first place.
We start from the equations of motion (\ref{DGEOM}), which we now view as an implicit
definition for \be
 \phi_{DG}=\phi_{DG}(\phi_X) ~. \label{phiDG-soln} \ee In the context of perturbation theory
it suffices to solve for $\phi_{DG}$ perturbatively. This can always be done given that
$L_{DG}$ is invertible (and local in 1d) as we proceed to detail.

One expands $\phi_{DG}$ as \be
 \phi_{DG} = \phi_{DG}^{(1)}(\phi_X) + \phi_{DG}^{(2)}(\phi_X) + \dots \ee
 where the superscript denotes the order with respect to the fields $\phi_X$. Next one substitutes this
expansion into the equations (\ref{DGEOM}) and solves them order
by order. At each order one obtains an equation of the form
$0=L_{DG}\, \phi_{DG}^{(k)} + Src^{(k)}$ where the $k$'th order
source term depends only upon $\phi_X$ as well as lower order of
$\phi_{DG}$. Now $\phi_{DG}^{(k)}$ can be solved for by the
assumed invertibility of $L_{DG}$.

The last step is to substitute (\ref{phiDG-soln}) into the action.

Altogether we have arrived at the canonical form of the action and
 completed our description of the process.

\section{Application to the non-uniform string}
\label{apply-sect}

In this section we spell out in greater detail how the general
procedure applies to the case of the non-uniform string.

The goals for such a computation include \bi
 \item The order of the GL phase transition requires expanding the
 action to fourth order, and was preformed in several works
 \cite{Gubser,SorkinD*,LG-GL} the last one employing the
 Landau-Ginzburg approach. The current method can be applied to
 further improve on the method by allowing to use a single
 ``master'' field, rather than a collection of three fields.

\item Performing computations along the non-uniform string branch
to a higher order than previously achieved. Such computations
could be compared to the numerical results of
\cite{Wiseman1,SorkinNUBS}. If one is
interested only in thermodynamics then it is necessary to compute
the action up to sixth order to go beyond the existing
literature. On the other hand, if one is interested in the
solutions themselves, only part of the third order was computed so
far.

\item Improving and extending the results on charged non-uniform strings
\cite{KudohMiyamoto1,KudohMiyamoto2}. However, in this section we
shall restrict ourselves to the neutral case for concreteness.
 \ei

\subsection*{Order of GL phase transition}

Following is the improved procedure to compute the order of the GL
phase transition at various space-time dimensions $d$.

\presub \emph{Set-up and identification of the DG fields}. The
space time coordinates are the standard $t,\, r,\, z$ and
$\Omega_{d-2}$. $t$ runs along the Euclidean time dimension which
has length $\beta$ asymptotically, $r$ is the radial coordinate,
$z$ runs along the compact dimension whose asymptotic size is $L$,
and $\Omega_{d-2}$ represents the angular coordinates on the $d-2$
sphere. The total space-time dimension is $d+1$. The gauge
functions obeying the isometries are $\xi_r,\, \xi_z$ (where
$\xi_\mu$ is the parameter of an infinitesimal diffeomorphism)
while the DG fields are those which are a vector or tensor from
the $r$ coordinate point of view, namely $h_{rr}$ and $h_{rz}$
(where $h_{\mu\nu}$ is the perturbation to the metric).

\presub \emph{Ansatz}. We need an ansatz which keeps the DG fields
and it is convenient to take the following constraint-free gauge
choice \be
 ds^2 = f(r)\, dt^2 + f(r)^{-1}\, e^{2b}\, dr^2 + e^{2\bt(r)}\, \( dz-\al\, dr
 \)^2 + r^2\, e^{2c}\, d\Omega_{d-2}^2 ~,\ee
where $b,\, \al,\, c$ are functions of $(r,z)$, $\bt$ is a
function of $r$ alone
%\footnote{$h$ was denoted by $\beta$ in \cite{nGPY}, but here we shall reserve the latter notation for the
% inverse temperature, and use $h$ instead as in \cite{LG-GL}.}
and $f(r)=1-(r_0/r)^{d-3}$. The uniform black string is obtained
upon setting all the fields to zero. In fixing the gauge we note
that for $z$-dependent modes we have two gauge functions, namely
$\xi_r,\, \xi_z$ which we can use to eliminate two fields chosen
here to be $h_{tt}$ and $h_{zz}$. However, for the zero mode
fields the situation is somewhat different \cite{nGPY}. As $\xi_z$
and $\al$ do not have a zero mode consistent with the symmetries
only $\xi_r$ can be used to eliminate a single field, chosen here
to be $h_{tt}$ and hence we retain the zero mode of $\bt$ (or
equivalently $h_{zz}$).

\presub \emph{Compute the $d+1$ dimensional action}. Compute the
Gibbons Hawking action. This could be achieved by computing the
Einstein-Hilbert action and integrating by parts to avoid terms
which include two $r$-derivatives acting on a single field. For
the purpose of computing the order of the transition it is enough
to retain terms up to fourth order in the fields according to the
Landau-Ginzburg theory.

\presub \emph{Reduction of the action to 1d}. Only the $z$
integration is non-trivial and we need to expand the fields in
harmonic functions. The action has two symmetries which constrain
the expansion. The first is $z \to -z$, and the second is $z \to z
+ L/2$ together with $\eps \to -\eps$, where $\eps$ is the small
expansion parameter along the non-uniform branch. Both symmetries
are remnants of the $z$-translation symmetry of the uniform
string. Expand \bea
 b(r,z) &=& \eps\, b_1\, \cos(k z) + \eps^2 \, b_0 + \eps^2\, b_2\, \cos(2 k z)
 \non
 \al(r,z) &=& \eps\, \al_1\, \sin(k z) +  \eps^2\, \al_2\, \sin(2 k z)\eea
 and $c$ is expanded just like $b$. All fields are now functions
of $r$ alone. Expand the action up to the fourth order in $\eps$
and perform all $z$ integrations.

\presub \emph{Integrate out the DG fields}. The DG fields are the
harmonics of $b,\,\al$ namely the five 1d fields
$b_1,b_0,b_2,\al_1,\al_2$. For the purpose of this application it
is not required to perturbatively solve an implicit equation. Next
substitute the solutions for the $DG$ fields back into the action.
Finally,  terms with a single derivative can be eliminated in some
circumstances through integration by parts.

\presub \emph{A Test}. The expressions for the quadratic action
should agree with those derived in \cite{nGPY}, namely \bea
 S_2 &=& S_2[c_1] + S_2[c_2] + S_2[\bt] + S_2[c_0] \non
 S_2[c_1] &=& \frac{(d-1)(d-2)}{2}\,  L\, \int dt\, r^{d-2} dr\,
 \[ f\, c_1'\,^2 - k^2 c_1^2 + V(r)\, c_1^2\] \non
 V(r) &=& -\frac{2(d-1)(d-3)^3}{r^2 \(2(d-2)r^{d-3}-(d-1)\)^2} \non
 S_2[\bt] &=& \frac{d-1}{d-2}\,  L\, \int dt\, f^2\, r^{d-2} dr\,
 \bt'\,^2 ~.
 \eea
The expression for $S_2[c_2]$ is gotten from that of $S_2[c_1]$ by
replacing $k \to 2 k$ while the expression for $S_2[c_0]$ can be
deduced from eq. (4.13) in \cite{nGPY}.

\presub \emph{ First order}. Write the equation of motion for
$c_1$ and solve the corresponding eigenvalue equation to obtain
$c_1 \equiv \phi_{GL}(r)$ and $k_0^2 \equiv \lambda_{GPY}$.

\presub \emph{ Second order}. Solve the back-reaction equations of
motion for the fields $c_2,\, c_0,\, h$.

\begin{figure}[t!] \centering \noindent
\includegraphics[width=7cm,angle=-90]{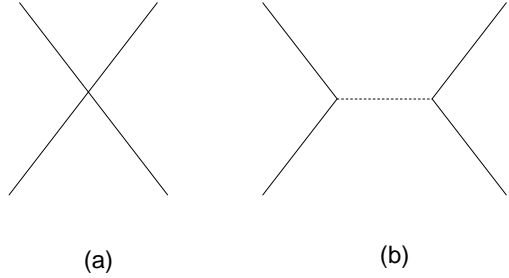}
\caption[]{The expansion of the action up to fourth order in terms
of Feynman diagrams. The solid four external legs represent the
$c_1$ field and the dashed internal propagator in (b) represents
all possible fields, namely $c_0,c_2,\bt$. (a) is the quartic
vertex of the GL mode while (b) describes the contribution from
the back-reaction fields.}\label{fig1}
\end{figure}

\presub \emph{Substitution in the action}. Compute $S_4(GL)$ and
$S_2(BR)$ - see \cite{LG-GL} for the definitions of these
notations. Continue to compute ${\cal C}=S_4(GL)-S_2(BR)$ which
determines the order of the phase transition in the canonical
ensemble (see also \cite{KudohMiyamoto1}). This computation is
encoded by the Feynman diagrams of figure \ref{fig1}. The results
can be compared against table 2 of \cite{LG-GL}. It is possible to
compute also the coefficient ${\cal A}$, allowing full
thermodynamic information including to pass to the micro-canonical
ensemble as in \cite{LG-GL}.

\subsection*{Higher orders}

The more efficient expressions gotten here should facilitate the
computation of higher orders. The perturbation theory can be
conveniently organized by means of Feynman diagrams. For instance,
in order to expand the action (or equivalently the free energy and
obtain the thermodynamics) up to order $j$ we need to compute all
tree diagrams (since it is a classical field theory) with $j$
external $c_1$ legs, and substitute $c_1 \equiv \phi_{GL}(r)$. In
this way we compute the free energy $F=F(\eps)$. Canonically the
free energy is a function of temperature, or equivalently of $k$
in our case. To obtain $F(k)$ we should compute also
$k=k(\eps)=k^{(0)} + \eps^2\, k^{(2)} + \dots$ to the required
order.

\subsection*{Acknowledgements}

This research is supported by The Israel Science Foundation grant no
607/05,  DIP grant H.52, EU grant MRTN-CT-2004-512194 and the
Einstein Center at the Hebrew University.

\end{document}